\documentclass[11pt,american]{article}
\usepackage[T1]{fontenc}
\usepackage[latin9]{inputenc}
\usepackage{geometry}
\usepackage{babel}
\usepackage{mathrsfs}
\usepackage{mathtools}
\usepackage{amsmath}
\usepackage{amssymb}
\usepackage[unicode=true]
 {hyperref}

\makeatletter
\newcommand{\lyxaddress}[1]{
\par {\raggedright #1
\vspace{1.4em}
\noindent\par}
}

\@ifundefined{date}{}{\date{}}
\usepackage{cite}
\usepackage{upgreek}

\makeatother

\begin{document}

\title{\textbf{Emergence of the Partial Trace}\\
\textbf{from Classical Probability Theory}}

\author{Andr\'es Macho-Ortiz,\textsuperscript{a,$\ast$} Francisco Javier
Fraile-Peláez,\textsuperscript{b} and Jos\'e Capmany\textsuperscript{a,c} }
\maketitle

\lyxaddress{{\small{}}\textsuperscript{{\small{}a}}{\small{}ITEAM Research
Institute, Universitat Polit\`ecnica de Val\`encia, Valencia 46022,
Spain}\\
{\small{}}\textsuperscript{{\small{}b}}{\small{}Dept. Teoría de
la Señal y Comunicaciones, Universidad de Vigo E.I. Telecomunicación,
Campus Universitario, E-36202 Vigo (Pontevedra), Spain}\\
{\small{}}\textsuperscript{{\small{}c}}{\small{}iPronics, Programmable
Photonics, S.L, Camino de Vera s/n, Valencia 46022, Spain}}

\noindent {\small{}$\ast$corresponding author: \href{mailto:amachor@iteam.upv.es}{amachor@iteam.upv.es}}{\small \par}

\vspace{1cm}
\begin{abstract}
The partial trace is commonly introduced in quantum mechanics as an
algebraic\linebreak{}
operation used to define reduced states of composite systems. However,
its connection with the classical rule of probabilistic marginalization
goes systematically unnoticed in the literature. Here, we show that
the partial trace emerges naturally from the requirement of consistency
between the Born rule for measurement probabilities and classical
marginalization. Starting from the relation between joint and marginal
probability distributions, we impose that the reduced density operator
of a subsystem reproduce the local measurement statistics derived
from the global state. We demonstrate that this requirement directly
leads to the standard expression of the partial trace. This derivation
reveals the reduced density operator as the quantum counterpart of
a marginal probability distribution and the partial trace as the corresponding
marginalization operation within the tensor-product framework of quantum
mechanics.\\
\end{abstract}
\noindent \begin{center}
\newpage{}
\par\end{center}

\section{Introduction}

\noindent Composite quantum systems play a central role in quantum
mechanics and quantum information theory. When a physical system is
composed of multiple subsystems, its state is described by a density
operator acting on the tensor-product Hilbert space of the individual
components \cite{key-1,key-2}. In many situations, however, one is
interested only in the state of a subsystem. The standard way to describe
such a situation is through the reduced density operator, obtained
by performing the partial trace over the degrees of freedom of the
complementary subsystem \cite{key-3,key-4,key-5,key-6,key-7}.

In the quantum-mechanical literature, the partial trace is typically
introduced as an\linebreak{}
algebraic operation acting on the density operator describing the
global quantum state, whose output is the reduced density operator
accounting for the local quantum state of a specific subsystem. Standard
textbooks define such reduced density operator as the operator that
reproduces the expectation values \cite{key-3,key-4,key-5,key-6}
and the measurement probabilities \cite{key-7} of any observable
acting locally on that subsystem, thus ensuring statistical consistency
between local observations and the global quantum state.

Although the mathematical definition is clear, it often leaves the
impression that the partial trace is a somewhat ad hoc algebraic operation.
However, the classical probabilistic origin underlying the definition
of the partial trace is rarely emphasized explicitly. In classical
probability theory, when dealing with joint probability distributions,
the probabilities of a subsystem are obtained through marginalization
over the variables of the complementary subsystem \cite{key-8,key-9}.
Given that the Born rule assigns probabilities to measurement outcomes
in quantum mechanics \cite{key-1,key-2}, it is natural to expect
that the reduced density operator should emerge from the combination
of the postulates of quantum mechanics and the classical marginalization
rule.

The aim of this work is to make this connection explicit. We show
that the standard expression of the partial trace can be derived directly
from the requirement that the reduced density operator reproduces
the marginal measurement probabilities associated with a composite
quantum system. Starting from the classical marginalization rule and
combining it with the Born rule for measurement probabilities, we
demonstrate that the reduced density operator must take the familiar
form defined by the partial trace. From this viewpoint, the partial
trace appears not as an independent algebraic definition but as a
natural consequence of the probabilistic framework of quantum mechanics.

\section{Preliminary mathematical concepts }

\noindent Consider two quantum systems $A$ and $B$, with state spaces
$\mathscr{E}_{A}$ and $\mathscr{E}_{B}$, respectively. Let $\widehat{\mathrm{A}}$
and $\widehat{\mathrm{B}}$ be two observables acting on $\mathscr{E}_{A}$
and $\mathscr{E}_{B}$, respectively. For clarity, we restrict our
attention to the case of non-degenerate discrete spectra (more general
situations, such as degenerate or continuous spectra, will be examined
in the Discussion section). The corresponding eigenvalue equations
describing the spectra of $\widehat{\mathrm{A}}$ and $\widehat{\mathrm{B}}$
are:
\begin{equation}
\widehat{\mathrm{A}}\bigr|a_{i}\bigr\rangle=a_{i}\bigr|a_{i}\bigr\rangle;\ \ \ \widehat{\mathrm{B}}\bigr|b_{j}\bigr\rangle=b_{j}\bigr|b_{j}\bigr\rangle;\ \ \ i,j\in\mathbb{N};\label{eq:2.1}
\end{equation}
with $\mathscr{E}_{A}$ and $\mathscr{E}_{B}$ being respectively
spanned by the states $\bigr|a_{i}\bigr\rangle$ and $\bigr|b_{j}\bigr\rangle$.
Here, when considering the composite system $AB$, the corresponding
state space is defined as $\mathscr{E}_{AB}\coloneqq\mathscr{E}_{A}\otimes\mathscr{E}_{B}$,
where $\otimes$ is the tensor product. Specifically, $\mathscr{E}_{AB}$
is spanned by the product states $\bigr|a_{i}\bigr\rangle\otimes\bigr|b_{j}\bigr\rangle$,
which are usually denoted as $\bigr|a_{i},b_{j}\bigr\rangle$ for
simplicity \cite{key-1,key-2}. In this scenario, it is convenient
to introduce the concept of ``extension operators'' to operate in
the composite system $AB$. In particular, the extension operators
of the observables $\widehat{\mathrm{A}}$ and $\widehat{\mathrm{B}}$
are respectively defined as $\widetilde{\mathrm{A}}\coloneqq\widehat{\mathrm{A}}\otimes\widehat{1}_{B}$
and $\widetilde{\mathrm{B}}\coloneqq\widehat{1}_{A}\otimes\widehat{\mathrm{B}}$,
with $\widehat{1}_{A}$ and $\widehat{1}_{B}$ being the identity
operators acting respectively on $\mathscr{E}_{A}$ and $\mathscr{E}_{B}$
\cite{key-1}. These extension operators, also termed extended observables,
are linear maps ($\mathscr{L}$) acting on $\mathscr{E}_{AB}$, explicitly
denoted as $\widetilde{\mathrm{A}},\widetilde{\mathrm{B}}\in\mathscr{L}\left(\mathscr{E}_{AB},\mathscr{E}_{AB}\right)$.

In quantum information, it is also useful to define linear maps that
combine operators with kets and bras through the tensor product, thereby
connecting the individual spaces $\mathscr{E}_{A}$ and $\mathscr{E}_{B}$
with the composite space $\mathscr{E}_{AB}$. Let $\left|a\right\rangle \in\mathscr{E}_{A}$
and $\left|b\right\rangle \in\mathscr{E}_{B}$ be arbitrary states.
Given a specific state $\left|b_{k}\right\rangle $ of $\mathscr{E}_{B}$,
we define the linear maps $\widehat{\mathrm{A}}\otimes\left|b_{k}\right\rangle \in\mathscr{L}\left(\mathscr{E}_{A},\mathscr{E}_{AB}\right)$
and $\widehat{\mathrm{A}}\otimes\left\langle b_{k}\right|\in\mathscr{L}\left(\mathscr{E}_{AB},\mathscr{E}_{A}\right)$
through the following actions \cite{key-7}:
\begin{align}
\left(\widehat{\mathrm{A}}\otimes\left|b_{k}\right\rangle \right)\left|a\right\rangle  & \coloneqq\widehat{\mathrm{A}}\left|a\right\rangle \otimes\left|b_{k}\right\rangle ,\label{eq:2.2}\\
\left(\widehat{\mathrm{A}}\otimes\left\langle b_{k}\right|\right)\left|a,b\right\rangle  & \coloneqq\widehat{\mathrm{A}}\left|a\right\rangle \left\langle b_{k}|b\right\rangle =\left\langle b_{k}|b\right\rangle \widehat{\mathrm{A}}\left|a\right\rangle .\label{eq:2.3}
\end{align}
These definitions extend the notion of extension operators to constructions
involving kets and bras. In particular, they allow us to rewrite,
for example, $\widehat{1}_{A}\otimes\bigr|b_{k}\bigr\rangle\bigl\langle b_{k}\bigr|$
as the composition of the maps $\bigl(\widehat{1}_{A}\otimes\bigr|b_{k}\bigr\rangle\bigr)\cdot\bigl(\widehat{1}_{A}\otimes\bigl\langle b_{k}\bigr|\bigr)$,
since both expressions induce the same mapping on the composite space
$\mathscr{E}_{AB}$:
\begin{align}
\left(\widehat{1}_{A}\otimes\bigr|b_{k}\bigr\rangle\bigl\langle b_{k}\bigr|\right)\left|a,b\right\rangle  & =\left(\widehat{1}_{A}\left|a\right\rangle \right)\otimes\left(\bigr|b_{k}\bigr\rangle\bigl\langle b_{k}\bigr|b\bigr\rangle\right)\nonumber \\
 & \equiv\bigl\langle b_{k}\bigr|b\bigr\rangle\left|a,b_{k}\right\rangle ,\label{eq:2.4}\\
\left(\widehat{1}_{A}\otimes\bigr|b_{k}\bigr\rangle\right)\cdot\left(\widehat{1}_{A}\otimes\bigl\langle b_{k}\bigr|\right)\left|a,b\right\rangle  & =\left(\widehat{1}_{A}\otimes\bigr|b_{k}\bigr\rangle\right)\left(\left|a\right\rangle \bigl\langle b_{k}\bigr|b\bigr\rangle\right)\nonumber \\
 & =\bigl\langle b_{k}\bigr|b\bigr\rangle\left(\widehat{1}_{A}\otimes\bigr|b_{k}\bigr\rangle\right)\left|a\right\rangle \nonumber \\
 & =\bigl\langle b_{k}\bigr|b\bigr\rangle\left|a,b_{k}\right\rangle .\label{eq:2.5}
\end{align}
Remarkably, the definitions provided by Eqs.\,(\ref{eq:2.2}) and
(\ref{eq:2.3}) will enable the intuitive derivation of the partial
trace from classical probability theory.

\section{Derivation of the partial trace from marginal probabilities}

\noindent In classical probability theory, the probability mass function
(pmf) of a random variable $A$ can be obtained from the joint pmf
of the bivariate random variable $\left(A,B\right)$ through marginalization
over the random variable $B$ \cite{key-8,key-9}:
\begin{equation}
P_{A}\left(a_{i}\right)=\sum_{j}P_{AB}\left(a_{i},b_{j}\right).\label{eq:3.1}
\end{equation}

In quantum mechanics, $P_{A}\left(a_{i}\right)$ denotes the probability
of obtaining the eigenvalue $a_{i}$ when measuring the observable
$\widehat{\mathrm{A}}$ on subsystem $A$. For a non-degenerate discrete
spectrum, this probability is given by the Born rule \cite{key-1,key-3}:
\begin{equation}
P_{A}\left(a_{i}\right)=\bigl\langle a_{i}\bigr|\widehat{\rho}_{A}\bigr|a_{i}\bigr\rangle,\label{eq:3.2}
\end{equation}
where $\widehat{\rho}_{A}$ denotes the (reduced) state describing
subsystem $A$. In other words, $\widehat{\rho}_{A}$ must include
in its populations $\bigl\langle a_{i}\bigr|\widehat{\rho}_{A}\bigr|a_{i}\bigr\rangle$
the marginal measurement probabilities $P_{A}\left(a_{i}\right)$
associated with the local observable $\widehat{\mathrm{A}}$. Since
$\widehat{\mathrm{A}}$ is arbitrary, this requirement extends to
any local observable acting on $\mathscr{E}_{A}$. Likewise, $P_{AB}\left(a_{i},b_{j}\right)$
denotes the joint probability of obtaining the outcome pair $\left(a_{i},b_{j}\right)$
when the extended observables $\widetilde{\mathrm{A}}=\widehat{\mathrm{A}}\otimes\widehat{1}_{B}$
and $\widetilde{\mathrm{B}}=\widehat{1}_{A}\otimes\widehat{\mathrm{B}}$
are measured on the composite system $AB$. Since $\widehat{\mathrm{A}}$
and $\widehat{\mathrm{B}}$ act on different subsystems, then $\widetilde{\mathrm{A}}$
and $\widetilde{\mathrm{B}}$ are compatible ($\bigl[\widetilde{\mathrm{A}},\widetilde{\mathrm{B}}\bigr]=0$),
and therefore admit a joint probability distribution \cite{key-1,key-2}.
The Born rule then gives:
\begin{equation}
P_{AB}\left(a_{i},b_{j}\right)=\bigl\langle a_{i},b_{j}\bigr|\widehat{\rho}_{AB}\bigr|a_{i},b_{j}\bigr\rangle,\label{eq:3.3}
\end{equation}
where $\widehat{\rho}_{AB}$ is the state of the composite system
$AB$. Hence, the classical marginalization rule Eq.\,(\ref{eq:3.1})
translates into the condition:
\begin{equation}
\bigl\langle a_{i}\bigr|\widehat{\rho}_{A}\bigr|a_{i}\bigr\rangle=\sum_{j}\bigl\langle a_{i},b_{j}\bigr|\widehat{\rho}_{AB}\bigr|a_{i},b_{j}\bigr\rangle.\label{eq:3.4}
\end{equation}

The key observation is that Eq.\,(\ref{eq:3.4}) is the direct quantum
analogue of the classical marginalization rule. We now show that imposing
this condition uniquely determines the reduced density operator $\widehat{\rho}_{A}$
of $\mathscr{E}_{A}$ and leads directly to the standard expression
of the partial trace:
\begin{equation}
\widehat{\rho}_{A}=\sum_{j}\bigl(\widehat{1}_{A}\otimes\left\langle b_{j}\right|\bigr)\widehat{\rho}_{AB}\bigl(\widehat{1}_{A}\otimes\left|b_{j}\right\rangle \bigr).\label{eq:3.5}
\end{equation}
This expression corresponds to the partial trace over subsystem $B$,
commonly written in the literature as \cite{key-3,key-4,key-5,key-6,key-7}:
\begin{equation}
\widehat{\rho}_{A}=\mathrm{Tr}_{B}\left(\widehat{\rho}_{AB}\right)\equiv\sum_{j}\bigl\langle b_{j}\bigr|\widehat{\rho}_{AB}\bigr|b_{j}\bigr\rangle,\label{eq:3.6}
\end{equation}
where $\equiv$ denotes a notational contraction that omits the identity
operator $\widehat{1}_{A}$. In this way, we will prove that the partial
trace naturally arises from the interplay between the postulates of
quantum mechanics and the classical marginalization rule.

\paragraph{Proof.}

We start from Eq.\,(\ref{eq:3.4}):
\begin{equation}
\bigl\langle a_{i}\bigr|\widehat{\rho}_{A}\bigr|a_{i}\bigr\rangle=\sum_{j}\bigl\langle a_{i},b_{j}\bigr|\widehat{\rho}_{AB}\bigr|a_{i},b_{j}\bigr\rangle,\label{eq:3.7}
\end{equation}
which reproduces the classical marginalization rule given by Eq.\,(\ref{eq:3.1})
in quantum mechanics. Here, we introduce the closure relation $\widehat{1}_{B}=\sum_{k}\bigr|b_{k}\bigr\rangle\bigl\langle b_{k}\bigr|$:
\begin{align}
\bigl\langle a_{i}\bigr|\widehat{\rho}_{A}\bigr|a_{i}\bigr\rangle & =\sum_{j}\bigl\langle a_{i},b_{j}\bigr|\left(\widehat{1}_{A}\otimes\widehat{1}_{B}\right)\widehat{\rho}_{AB}\left(\widehat{1}_{A}\otimes\widehat{1}_{B}\right)\bigr|a_{i},b_{j}\bigr\rangle\nonumber \\
 & =\sum_{j,k,k^{\prime}}\bigl\langle a_{i},b_{j}\bigr|\left(\widehat{1}_{A}\otimes\bigr|b_{k}\bigr\rangle\bigl\langle b_{k}\bigr|\right)\widehat{\rho}_{AB}\left(\widehat{1}_{A}\otimes\bigr|b_{k^{\prime}}\bigr\rangle\bigl\langle b_{k^{\prime}}\bigr|\right)\bigr|a_{i},b_{j}\bigr\rangle.\label{eq:3.8}
\end{align}
Next, we apply the identity demonstrated in Eqs.\,(\ref{eq:2.4})
and (\ref{eq:2.5}):
\begin{equation}
\widehat{1}_{A}\otimes\bigr|b_{k}\bigr\rangle\bigl\langle b_{k}\bigr|=\left(\widehat{1}_{A}\otimes\bigr|b_{k}\bigr\rangle\right)\cdot\left(\widehat{1}_{A}\otimes\bigl\langle b_{k}\bigr|\right),\label{eq:3.9}
\end{equation}
and we omit the dot ``$\cdot$'' to simplify the notation. Hence,
we have:
\begin{align}
\bigl\langle a_{i}\bigr|\widehat{\rho}_{A}\bigr|a_{i}\bigr\rangle & =\sum_{j,k,k^{\prime}}\bigl\langle a_{i},b_{j}\bigr|\left(\widehat{1}_{A}\otimes\bigr|b_{k}\bigr\rangle\right)\left(\widehat{1}_{A}\otimes\bigl\langle b_{k}\bigr|\right)\widehat{\rho}_{AB}\left(\widehat{1}_{A}\otimes\bigr|b_{k^{\prime}}\bigr\rangle\right)\left(\widehat{1}_{A}\otimes\bigl\langle b_{k^{\prime}}\bigr|\right)\bigr|a_{i},b_{j}\bigr\rangle.\label{eq:3.10}
\end{align}
Then, using the identities:
\begin{align}
\left(\widehat{1}_{A}\otimes\bigl\langle b_{k^{\prime}}\bigr|\right)\bigr|a_{i},b_{j}\bigr\rangle & =\bigr|a_{i}\bigr\rangle\bigl\langle b_{k^{\prime}}\bigr|b_{j}\bigr\rangle\equiv\bigr|a_{i}\bigr\rangle\delta_{k^{\prime},j},\label{eq:3.11}\\
\bigl\langle a_{i},b_{j}\bigr|\left(\widehat{1}_{A}\otimes\bigr|b_{k}\bigr\rangle\right) & =\bigl\langle b_{j}\bigr|b_{k}\bigr\rangle\bigl\langle a_{i}\bigr|\equiv\delta_{jk}\bigl\langle a_{i}\bigr|,\label{eq:3.12}
\end{align}
we obtain:
\begin{align}
\bigl\langle a_{i}\bigr|\widehat{\rho}_{A}\bigr|a_{i}\bigr\rangle & =\sum_{j,k,k^{\prime}}\delta_{jk}\bigl\langle a_{i}\bigr|\left(\widehat{1}_{A}\otimes\bigl\langle b_{k}\bigr|\right)\widehat{\rho}_{AB}\left(\widehat{1}_{A}\otimes\bigr|b_{k^{\prime}}\bigr\rangle\right)\bigr|a_{i}\bigr\rangle\delta_{k^{\prime},j}\nonumber \\
 & =\sum_{j}\bigl\langle a_{i}\bigr|\left(\widehat{1}_{A}\otimes\bigl\langle b_{j}\bigr|\right)\widehat{\rho}_{AB}\left(\widehat{1}_{A}\otimes\bigr|b_{j}\bigr\rangle\right)\bigr|a_{i}\bigr\rangle\nonumber \\
 & =\bigl\langle a_{i}\bigr|\sum_{j}\left(\widehat{1}_{A}\otimes\bigl\langle b_{j}\bigr|\right)\widehat{\rho}_{AB}\left(\widehat{1}_{A}\otimes\bigr|b_{j}\bigr\rangle\right)\bigr|a_{i}\bigr\rangle.\label{eq:3.13}
\end{align}
This identity holds for any ket $\bigr|a_{i}\bigr\rangle$ of the
vector basis $\bigl\{\bigr|a_{i}\bigr\rangle\bigr\}_{i}$ and for
any state $\widehat{\rho}_{A}$ of $\mathscr{E}_{A}$ and $\widehat{\rho}_{AB}$
of $\mathscr{E}_{AB}$. Therefore, we can conclude that:
\begin{equation}
\widehat{\rho}_{A}=\sum_{j}\left(\widehat{1}_{A}\otimes\bigl\langle b_{j}\bigr|\right)\widehat{\rho}_{AB}\left(\widehat{1}_{A}\otimes\bigr|b_{j}\bigr\rangle\right)\equiv\sum_{j}\bigl\langle b_{j}\bigr|\widehat{\rho}_{AB}\bigr|b_{j}\bigr\rangle,\label{eq:3.14}
\end{equation}
with $\equiv$ accounting for a symbolic contraction that omits the
identity operator $\widehat{1}_{A}$. The above equation corresponds
to the well-known expression of the partial trace over subsystem $B$,
usually denoted as $\widehat{\rho}_{A}=\mathrm{Tr}_{B}\left(\widehat{\rho}_{AB}\right)$
(see e.g. \cite{key-3,key-4}).

\section{Discussion}

\noindent The derivation presented in this work provides a probabilistic
interpretation of the partial trace that complements the standard
algebraic treatments commonly found in the literature. In most textbooks
on quantum mechanics and quantum information theory, the reduced density
operator is introduced as the operator that reproduces the measurement
statistics of all observables acting locally on a subsystem. Within
this framework, the partial trace is introduced primarily as a mathematical
operation that ensures the consistency between the global state of
a composite system and the statistics of local measurements \cite{key-3,key-4,key-5,key-6,key-7}. 

Although this definition is mathematically precise, it leaves unclear
its classical probabilistic origin. In particular, the close connection
with the classical marginalization rule is not emphasized explicitly.
In classical probability theory, the probability distribution of a
subsystem is obtained from the joint distribution of a composite system
by summing over the outcomes of the complementary subsystem \cite{key-8,key-9}.
Since the Born rule assigns probabilities to measurement outcomes
in quantum mechanics \cite{key-1}, it is natural to expect that the
description of subsystems should emerge from an analogous marginalization
principle. From this perspective, the reduced density operator may
be viewed as the quantum counterpart of a marginal probability distribution,
suggesting that the partial trace should arise as the corresponding
marginalization operation.

The derivation presented here makes this connection explicit. By imposing
that the reduced density operator must reproduce the marginal measurement
probabilities associated with the composite system {[}Eq.\,(\ref{eq:3.4}){]},
we showed that the operator describing subsystem $A$ must take the
form $\widehat{\rho}_{A}=\mathrm{Tr}_{B}\left(\widehat{\rho}_{AB}\right)=\sum_{j}\bigl\langle b_{j}\bigr|\widehat{\rho}_{AB}\bigr|b_{j}\bigr\rangle$,
which corresponds to Eq.\,(\ref{eq:3.14}). Thus, the standard expression
of the partial trace emerges directly from the requirement that quantum
measurement probabilities obey the same marginalization principle
that relates joint and marginal probability distributions in classical
probability theory.

Although we have restricted the discussion to non-degenerate discrete
spectra for simplicity, the derivation extends naturally to more general
situations, including degenerate and continuous spectra. For instance,
non-degenerate continuous spectra requires to start the derivation
from the classical marginalization rule relating marginal and joint
probability density functions. In such a case, Eq.\,(\ref{eq:3.1})
becomes \cite{key-8,key-9}:
\begin{equation}
f_{A}\left(a\right)=\int_{-\infty}^{\infty}f_{AB}\left(a,b\right)\mathrm{d}b.\label{eq:4.1}
\end{equation}
In quantum mechanics \cite{key-1}, $f_{A}$ denotes the probability
density associated with obtaining an outcome of the observable $\widehat{\mathrm{A}}$
within the interval $\left[a,a+\mathrm{d}a\right]$ when it is locally
measured on subsystem $A$. Likewise, $f_{AB}$ denotes the joint
probability density associated with obtaining an outcome of the extended
observables $\widetilde{\mathrm{A}}=\widehat{\mathrm{A}}\otimes\widehat{1}_{B}$
and $\widetilde{\mathrm{B}}=\widehat{1}_{A}\otimes\widehat{\mathrm{B}}$
within the region $\left[a,a+\mathrm{d}a\right]\times\left[b,b+\mathrm{d}b\right]$
when they are measured on the composite system $AB$. By rewriting
Eq.\,(\ref{eq:4.1}) in terms of the density operators $\widehat{\rho}_{A}$
and $\widehat{\rho}_{AB}$, following the same reasoning presented
in Section 3, the partial trace finally emerges of the form:
\begin{align}
\widehat{\rho}_{A} & =\mathrm{Tr}_{B}\left(\widehat{\rho}_{AB}\right)=\int_{-\infty}^{\infty}\left(\widehat{1}_{A}\otimes\bigl\langle b\bigr|\right)\widehat{\rho}_{AB}\left(\widehat{1}_{A}\otimes\bigr|b\bigr\rangle\right)\mathrm{d}b\equiv\int_{-\infty}^{\infty}\bigl\langle b\bigr|\widehat{\rho}_{AB}\bigr|b\bigr\rangle\mathrm{d}b.\label{eq:4.2}
\end{align}
The same argument can be extended straightforwardly to degenerate
discrete and continuous spectra.

From this perspective, the partial trace does not appear as an independent
algebraic\linebreak{}
definition but as a structural consequence of two fundamental ingredients:
the tensor-product description of composite quantum systems and the
probabilistic interpretation provided by the Born rule. This viewpoint
highlights the close relationship between the mathematical formalism
of quantum mechanics and the classical theory of probability. In particular,
it clarifies that the reduced density operator plays a role analogous
to that of a marginal probability distribution in classical statistics,
while the partial trace emerges as the corresponding marginalization
operation within the quantum formalism. 

Beyond providing an alternative conceptual interpretation, this viewpoint
also has pedagogical implications. Students encountering the formalism
of composite quantum systems often perceive the partial trace as a
technical algebraic construction. Interpreting it instead as the quantum
counterpart of classical marginalization, arising from the consistency
between the Born rule and classical probability theory, provides a
more intuitive route to the concept. This perspective helps bridge
the conceptual gap between classical probability theory and quantum
mechanics.

Finally, the probabilistic viewpoint adopted here complements other
foundational approaches that attempt to derive aspects of quantum
mechanics from information-theoretic or probabilistic principles.
For instance, Benavoli \emph{et al.} have shown that the formalism
of quantum mechanics can be interpreted as a generalization of Bayesian
probability theory to the space of Hermitian matrices, where operations
such as measurement, marginalization, and independence correspond
to probabilistic rules in this generalized framework \cite{key-10}.
While that approach provides a broad probabilistic reinterpretation
of quantum mechanics, it does not explicitly address the derivation
of the partial trace from the requirement of consistency between the
Born rule and classical marginalization. In \cite{key-10} the partial
trace is interpreted as the counterpart of classical marginalization
in the space of density operators. In contrast, the present work shows
that the standard expression of the partial trace follows directly
from imposing the classical marginalization principle on the probabilities
assigned by the Born rule. From this standpoint, the partial trace
naturally emerges as the quantum counterpart of classical marginalization,
preserving the probabilistic consistency of subsystems within the
tensor-product structure of quantum theory.

\subsection*{Acknowledgements}

\noindent This work was supported by ERC-ADG-2022-101097092 ANBIT,
ERC-POC-2025-1 101241773 TRANSBIT MESH, GVA PROMETEO 2021/015 research
excellency award, Fundación BBVA Programa de Investigación Fundamentos
2024 API project, Ministerio de Ciencia y Universidades Plan Complementario
de Comunicación Cuántica projects QUANTUMABLE-1 and QUANTUMABLE-2,
and HUB de Comunicaciones Cuánticas. The authors also thank Jaime
Gimeno Balaguer for fruitful discussions on the probabilistic interpretation
of the partial trace.


\begin{thebibliography}{10}
\bibitem[1]{key-1} C. C.-Tannoudji, B. Diu, and F. Laloë, \emph{Quantum
Mechanics, Volume I: Basic Concepts, Tools, and Applications}. (Wiley,
Weinheim, 2020).

\bibitem[2]{key-2} J. J. Sakurai, \emph{Modern Quantum Mechanics}.
(Addison-Wesley, 1994).

\bibitem[3]{key-3} S. M. Barnett, \emph{Quantum Information}. (Oxford
University Press, Oxford, 2009).

\bibitem[4]{key-4} M. A. Nielsen and I. L. Chuang, \emph{Quantum
Computation and Quantum Information}. (Cambridge University Press,
Cambridge, 2016).

\bibitem[5]{key-5} G. Benenti, G. Casati, and G. Strini, \emph{Principles
of Quantum Computation and Information. Volume II: Basic Tools and
Special Topics}. (World Scientific, London, 2007).

\bibitem[6]{key-6} B. Vacchini, \emph{Open Quantum Systems: Foundations
and Theory}. (Springer, Cham, 2024).

\bibitem[7]{key-7} M. M. Wilde, \emph{Quantum Information Theory}.
(Cambridge University Press, Cambridge, 2017).

\bibitem[8]{key-8} H. P. Hsu, \emph{Probability, Random Variables,}
\& \emph{Random Processes}. (McGraw-Hill, 1997).

\bibitem[9]{key-9} A. Papoulis, \emph{Probability, Random Variables,
and Stochastic Processes}. (McGraw-Hill, 2002).

\bibitem[10]{key-10} A. Benavoli, A. Facchini, and M. Zaffalon, ``Quantum
mechanics: the Bayesian theory generalized to the space of Hermitian
matrices,'' Physical Review A \textbf{94}, 042106 (2016).
\end{thebibliography}
\end{document}